\documentstyle[12pt,citesort]{article}
\def\IJMP #1 #2 #3 {{\it Int.\ J.\ Mod.\ Phys.}\ {\bf #1}\ (#2) #3}
\def\MPL #1 #2 #3 {{\it Mod.\ Phys.\ Lett.}\ {\bf #1}\ (#2) #3}
\def\NC #1 #2 #3 {{\it Nuovo Cim.}\ {\bf #1} (#2) #3}
\def\NP #1 #2 #3 {{\it Nucl.\ Phys.}\ {\bf #1}\ (#2) #3}
\def\PL #1 #2 #3 {{\it Phys.\ Lett.}\ {\bf #1}\ (#2) #3}
\def\PR #1 #2 #3 {{\it Phys.\ Rev.}\ {\bf #1}\ (#2) #3}
\def\PP #1 #2 #3 {{\it Phys.\ Rep.}\ {\bf #1}\ (#2) #3}
\def\PRL #1 #2 #3 {{\it Phys.\ Rev.\ Lett.}\ {\bf #1}\ (#2) #3}
\def\RMP #1 #2 #3 {{\it Rev.\ Mod.\ Phys.}\ {\bf #1}\ (#2) #3}
\def\ZP #1 #2 #3 {{\it Z.\ Phys.}\ {\bf #1}\ (#2) #3}
\def\g{\gamma}
\begin{document}
\begin{flushright}
hep-ph/9406428\\
June 1994
\end{flushright}
\vspace*{10mm}
\begin{center}
{\bf QCD CORRECTIONS TO $b\bar b/c\bar c$ PAIR PRODUCTION IN POLARIZED
$\g\g$ COLLISIONS AND THE INTERMEDIATE MASS HIGGS SIGNAL}\footnote{Based on
invited talk given at the {\it ``Workshop on gamma--gamma colliders"},
March~28-31, 1994, Lawrence Berkeley Laboratory}\\

\vspace*{5mm}
{G. Jikia and A.~Tkabladze}\\
[1ex]	{\it Institute for High Energy Physics} \\
	{\it 142284, Protvino, Moscow region, Russia} 
\end{center}

\begin{abstract}
We present production rates of the two- and three-jet final states for the
processes of massive $c\bar c/b\bar b$ quark production in circularly
polarized photon-photon collisions, including QCD radiative corrections.
Lowest order cross section, one-loop virtual correction and gluon emission
correction are shown to be of the same order of magnitude for $b\bar b$
quark production at $\sqrt{s_{\g\g}}\sim 100$~GeV. It is shown that the
signal from intermediate mass Higgs boson is observable at photon-photon
collider, although the statistical significance is substantially reduced
with respect to the tree level calculation.
\end{abstract}

A very exciting potential application of photon-photon collisions at a
100-200~GeV linear collider is the intermediate mass Higgs boson production
in photon fusion reaction \cite{gunion,borden,threejet,zerwas,brodsky}
$$\g\g\to H\to b\bar b.$$ In addition, new interesting method was proposed
\cite{grz,KKSZ} to measure the parity of the Higgs states in linearly
polarized photon-photon collisions.  It provides an opportunity to
investigate non-trivial assignment of the quantum numbers for Higgs
particles in extended models such as supersymmetric theories which include
both scalar $0^{++}$ and pseudoscalar $0^{-+}$ states \cite{SUSY}.

Extracting the intermediate mass Higgs signal in photon-photon collisions is
a hard task since large number of $b\bar b/c\bar c$ background events must
be rejected \cite{gunion,borden,eboli}. The crucial assumption is that these
large backgrounds can be actively suppressed by exploiting the polarization
dependence of the cross sections. Far above the threshold, the
$\g\g\to q\bar q$ cross section is dominated by initial photons in
the $J_z=\pm 2$ helicity state. Taking into account that the Higgs signal
comes from the $J_z=0$ channel, polarized collisions can be used to enhance
the signal while simultaneously suppressing the background
\cite{gunion,borden} (see also detailed discussion in these proceedings
\cite{procE,procB}). But the question remains how do QCD radiative
corrections influence these conclusions. While it is known that far above
the threshold the magnitude of these corrections is moderate for unpolarized
collisions
\cite{KMS,DKZZ}, one can expect that their effects will be especially large
for $q\bar q$ production in $J_z=0$ helicity state, where the tree level
contribution is suppressed as $m_q^2/s$. 

We present here the results of the one-loop calculation of the QCD
corrections to $b\bar b/c\bar c$ quark pair production in polarized
photon-photon collisions retaining the full dependence on the quark masses.
The total cross section calculated up to the order $\alpha^2\alpha_s$ is given
by the sum of tree-level $\g\g\to q\bar q$ contribution, the interference
term between one-loop and the tree level contributions, and tree level
contribution from quark pair production accompanied by gluon emission
$\g\g\to q\bar q g$. The first two contributions lead to two parton final
states converting mainly into two jets, while the third one leads to three
parton production converting both into two- and three-jet final states. The
reason is that three parton final states with collinear and/or soft gluon
will appear experimentally as two jets.  Moreover, only the sum of the cross
sections of $q\bar q$ production and $q\bar q g$ with soft or collinear
gluon is free from infrared divergences and has no singularities in the
limit $m_q\to 0$.  So, as usual, we consider the three parton state to
represent two-jet final state if the invariant mass of two partons is
sufficiently small $$ s_{i,j} < y_{cut} s_{\g\g}, $$ where
$s_{i,j}=(p_i+p_j)^2$ is the invariant mass squared of two partons $i$ and
$j$ and $\sqrt{s_{\g\g}}$ is the total c.m.s.  energy of two colliding
photons.

Fig.~1 shows total ({\it i.e.} two-jet plus three-jet) and two-jet
($y_{cut}=0.08$) cross sections for $b\bar b/c\bar c$ pair production in
polarized monochromatic $\g\g$ collisions. While the QCD corrections for
$J_z=\pm 2$ photon helicities are quite small, those for $J_z=0$ enhance
$c\bar c$ production by an order of magnitude or even larger. For $b\bar b$
production the situation is more complicated: the corrected total cross
section is smaller than the tree level $\g\g\to b\bar b$ cross section for
$\sqrt{s_{\g\g}}< 85$~GeV and larger for larger energies. The effect is more
pronounced for two-jet production. For small values of $y_{cut}<0.04$ the
two-jet differential cross section is even negative in some regions of the
phase space. This means that for $b\bar b$ production at
$\sqrt{s_{\g\g}}\sim 100$~GeV all three contributions (lowest order, virtual
and gluon emission) are of the same order of magnitude. This is unlike 
the case of $c\bar c$ production, where the gluon emission contribution
dominates. Therefore the approach of \cite{threejet,procB}, where only one
contribution from radiative processes $\g\g\to c\bar c g,\ b\bar b g$ is
taken into account in the limit $m_c$, $m_b\to 0$ (for $J_z=0$ and $m_q=0$
the cross section is given by this only contribution), may be relevant for
$c\bar c$ production, but is definitely not applicable for $b\bar b$
production.

Fig.~2 shows the events rates of signal and background two-jet final states
in photon-photon collisions. We make here the same assumptions as in
\cite{borden}, {\it i.e.} we choose the broad photon-photon luminosity
spectrum resulting from polarized linac and laser in the
$\lambda_{\g}\lambda_e>0$ direction, $\lambda_e=0.9$, $\lambda_{\gamma}=1$,
parameter $x=4.8$ and geometric factor $\rho = 0.6$ \cite{telnov,borden}. We
also assume the linac beam energy to be equal to 125~GeV and integrated
effective luminosity of 10~fb$^{-1}$. Such a choice is preferable when
trying to cover the entire intermediate Higgs mass region. We ignore here
backgrounds from $e\g\to eZ\to eb\bar b$ and $\g\g\to f\bar f Z$ processes
\cite{HZ}, which are essential for $m_H\sim m_Z$. The backgrounds coming
from the resolved photon contributions $\g g\to b\bar b,\ c\bar c$ are also
shown. While resolved photon contributions make almost impossible to observe
the intermediate mass Higgs signal at 500~GeV linear collider
\cite{eboli}, these backgrounds are much less significant at 250~GeV due to
a steeply falling gluon spectrum (see also \cite{borden,KKSZ}). QCD
corrections to Higgs decay into $b\bar b$ \cite{Hbb} are also taken into
account. We use a cut of $|\cos\theta|<0.7$ in the laboratory frame and not
in c.m. frame as in \cite{borden}. Cut in the laboratory frame gives
slightly better statistical significance of the Higgs signal. Finally, we
assume 5\%  $c\bar c$-to-$b\bar b$ misidentification probability. Thus, the
combined background ({\it i.e.} $b\bar b + 0.05 c\bar c$) is represented by
dotted line and can be compared with the signal denoted by solid line.

Fig.~3 presents the statistical significance of the Higgs boson signal
estimated from tree level as well as one-loop calculations including the
resolved photon contributions. This plot assumes  90\% $b\bar b$-tagging
efficiency for the $b\bar b$ final states and the resolution for
reconstructing the invariant mass of a two-jet events to be Gaussian with
$FWHM=0.1 m_H$. From this figure one can conclude that it is advantageous to
select two-jet final states and to impose the angular cut in the laboratory
frame. The account of QCD corrections reduces the statistical significance
of Higgs signal almost by a factor of two in comparison to tree-level
result. Nevertheless the intermediate mass Higgs boson can be observed in
$\g\g$ collisions at least at the level of 5$\sigma$ in the mass interval
from 80 to 160~GeV. Our estimates here should be considered as a first-order
determination of the influence of QCD radiative corrections on the
statistical error in the measurement of the two-photon Higgs width. A more
detailed analyses including full detector simulation is certainly needed.

\vspace*{1cm}

We are grateful to D.~Borden and O.~\' Eboli for helpful discussions.
Special thanks to organizers of the Workshop for financial help and to
A.~Sessler and M.~Chanowitz for kind hospitality. The attendance at the
Workshop was supported, in part, by the International Science Foundation
travel grant.

\newpage
\section*{Figure captions}
\parindent=0pt
\parskip=\baselineskip

Fig.~1. Cross sections for $\g\g\to c\bar c$ and $\g\g\to b\bar b$ for
polarized monochromatic photon beams, $++$, $+-$ correspond to $J_z=0$ and
$J_z=\pm 2$, respectively. Solid (dashed) line is tree level (one-loop
corrected) cross section for $b\bar b$ production. Dotted (dash-dotted) line
is tree level (one-loop corrected) cross section for $c\bar c$ production.
First figure gives total cross section. The second is two-jet production
cross section.

Fig.~2. Expected event rates for the Higgs signal and background processes.
Two-jet final states are considered. Lowest order as well as one-loop
corrected results are shown.

Fig.~3. Statistical significance of the intermediate mass Higgs boson
signal. Solid line corresponds to two-jet final states with the angular cut
in the laboratory frame. Dashed curves correspond to cut in c.m.s. of
colliding photons. ``All'' represents the sum of two- and three-jet final
states. Dash-dotted curves represent the results of the lowest order
calculation and the effect of the account of resolved photon contribution.

\end{document}